\documentclass[journal]{IEEEtran}
\ifCLASSINFOpdf
\usepackage[pdftex]{graphicx}
\else
\fi

\usepackage{braket}

\usepackage{cite}

\begin{document}
%
\title{A cost-effective measurement-device-independent quantum key distribution system for quantum networks}
%
%
%

\author{Raju Valivarthi, Qiang Zhou, Caleb John,  Francesco Marsili, Varun B. Verma, Matthew~D.~Shaw, Sae Woo Nam, Daniel Oblak, and  Wolfgang Tittel
\thanks{R.~Valivarthi, Q.~Zhou, D.~Oblak and W.~Tittel are with the Department of Physics and Astronomy, and the Institute for Quantum Science and Technology, University of Calgary, Calgary, T2N 1N4, Canada (e-mail: vrrvaliv@ucalgary.ca, qian.zhou@ucalgary.ca, doblak@ucalgary.ca and wtittel@ucalgary.ca).}
\thanks{C.~John is with the Department of Electrical and Computer Engineering, and the Institute for Quantum Science and Technology, University of Calgary, Calgary, AB, T2N 1N4, Canada (e-mail: ctjohn@ucalgary.ca).}
\thanks{F.~Marsili and M.~D.~Shaw are with the Jet Propulsion Laboratory, California Institute of Technology, Pasadena, CA, 91109, USA (e-mail: Francesco.Marsili.Dr@jpl.nasa.gov and Matthew.D.Shaw@jpl.nasa.gov).}
\thanks{V.~B.~Verma and S.~W.~Nam are with the National Institute of Standards and Technology, Boulder, CO, 80305, USA (e-mail: varun.verma@nist.gov and saewoo.nam@nist.gov).}}
\maketitle

\begin{abstract}
We experimentally realize a measurement-device-independent quantum key distribution (MDI-QKD) system based on cost-effective and commercially available hardware such as distributed feedback (DFB) lasers and field-programmable gate arrays (FPGA) that enable time-bin qubit preparation and time-tagging, and active feedback systems that allow for compensation of time-varying properties of photons after transmission through deployed fibre. We examine the performance of our system, and conclude that its design does not compromise performance. Our demonstration paves the way for MDI-QKD-based quantum networks in star-type topology that extend over more than 100 km distance.

\end{abstract}


%
\IEEEpeerreviewmaketitle

\section{Introduction}
%
%
%
%
\IEEEPARstart{B}{eing} the most mature quantum information technology, quantum key distribution (QKD) allows establishing cryptographic keys between two distant users (commonly known as Alice and Bob) based on the laws of quantum mechanics~\cite{BB84,Gisin2002,Scarani2009,Lo2014}. In conjunction with one-time-pad (OTP) encoding, QKD thereby provides a way for provably secure communication, thus promising to end the ongoing battle between codemakers and codebreakers. Many QKD systems, including commercial systems, have been developed during the last 30 years \cite{Gisin2002,Scarani2009,Lo2014,commercial}, and figures-of-merit such as secret key rates and maximum transmission distance continue to improve. However, quantum hacking over the past decade has also established that the specifications of components and devices used in actual QKD systems never perfectly agree with the theoretical description used in security proofs, which can compromise the security of real QKD systems. For instance, the so-called `blinding attacks' exploit vulnerabilities of single photon detectors (SPDs) to open a side-channel via which an eavesdropper can gain full information about the (assumed-to-be) secure key \cite{Lydersen2010}. Making practical QKD systems secure against all such attacks is a challenging task that has been investigated by many research groups. One approach is to develop attack-specific counter-measures \cite{Zhao2008,Yuan2010,Lim2014}. Unfortunately, the success of this strategy strongly depends on how well a QKD system is characterized, and the security of a `patched' QKD system may be compromised in future due to new and unforeseen attacks. A better solution is to devise and implement protocols that are intrinsically free of all side-channel attacks such as device-independent QKD (DI-QKD) \cite{Acin2007}, whose security is guaranteed by a loophole-free Bell test. Although such tests have recently been reported \cite{Hanson2015, Zeilinger2015, Nam2015}, the implementation of long distance DI-QKD still seems unlikely in the near future.

  Several groups, rather than eliminating the vulnerability to all side-channel attacks, have recently started to focus on QKD protocols  that are immune to the most dangerous side-channel attacks, i.e. all possible (known or yet-to-be proposed) detector side-channel attacks. One of these protocols is inspired by time-reversed entanglement-based QKD \cite{Biham1996, Inamori2002} and is known as measurement-device-independent QKD (MDI-QKD)~\cite{Lo2012}. It requires a Bell state measurement (BSM) at a central station, usually referred-to as Charlie, to create entanglement-like correlations between Alice and Bob. The key feature is that even if an eavesdropper completely controls the measurement devices (e.g. by replacing the BSM by another measurement), she would not be able to gain any information about the distributed key without Alice and Bob noticing. This means no assumptions are required about the measurement devices to guarantee the security of MDI-QKD, thus making it intrinsically immune to all detector side-channel attacks. Furthermore, due to the possibility for a large number of users to connect to the same Charlie, point-to-point MDI-QKD is ideally suited for extension into star-type networks. And last but not least, MDI-QKD can be seamlessly upgraded -- not disruptively replaced -- into quantum repeater-based long-distance quantum communication as more mature hardware becomes available \cite{Lvovsky2009, Sangouard2011}. 
  
  Due to the above-mentioned advantages, MDI-QKD has received much attention over the past 5 years and has meanwhile been demonstrated by several experimental groups in different configurations. Initial experiments consisted of proof-of-principle demonstrations in a real-world environment with time-bin qubits \cite{Wolfgang2013} and in the laboratory using the same encoding but with random selection of bases and states (as is required for secure key generation) \cite{Liu2013}. Subsequent demonstrations have included MDI-QKD in the laboratory with polarization qubits  \cite{daSilva2013,Tang2014}. Furthermore, long distance MDI-QKD has  been achieved over 200 km and 404 km of spooled optical fiber \cite{Tang200, Yin404}, and been mimicked using a short fibre with additional 60 dB loss \cite{Raju2015}. More recently, additional point-to-point and network field tests have been implemented \cite{Tang30,Tang_Net}.

In our previous studies, we have demonstrated a proof-of-principle of MDI-QKD over deployed fibre \cite{Wolfgang2013}, and also assessed the impact of using different single-photon detectors and different methods for time-bin qubit preparation on the performance of MDI-QKD~\cite{Raju2015}. The comprehensive understanding of trade-offs among complexity, cost, and system performance acquired through these investigations has now allowed us to develop a complete system, which is described and characterized in the following sections. In particular, our MDI-QKD system now also includes time tagging of qubit generations and detections, which allows key generation from qubits in randomly prepared states, and further improved polarization and arrival-time control of photons travelling from Alice and Bob to Charlie, which ensures their indistinguishability at the moment of the BSM. After examining the overall system performance (i.e. secret key rates and maximally-tolerable transmission loss), we conclude that our cost-effective implementation does not compromise the performance of MDI-QKD. We note that our real-time and continuously-running photon-state control can also be employed in other real-world systems that require stabilization of channels for quantum information transfer, e.g. in quantum teleportation \cite{Raju2016}.
\section{Protocol}
\label{sec:protocol}
In the MDI-QKD protocol, the two users -- Alice and Bob -- prepare qubits randomly in one of the four BB84 states. In the case of time-bin qubits, these are $\ket{e}$, $\ket{l}$, $\ket{+} \equiv (\ket{e} +\ket{l})/\sqrt{2}$, and $\ket{-} \equiv (\ket{e} -\ket{l})/\sqrt{2}$. Here, $\ket{e}$ and $\ket{l}$ denote the emission of a photon in an early and late temporal mode, respectively, forming the so-called Z-basis, while $\ket{+}$ and $\ket{-}$ describe superpositions of photon emissions and form the X-basis. Alice and Bob agree that $\ket{+}$ and $\ket{e}$ correspond to a classical bit value of `0', and $\ket{-}$ and $\ket{l}$ to `1'. The prepared qubits are sent to a third party, generally referred-to as Charlie. Charlie performs a BSM that projects the joint state of the two qubits (one from Alice and  one from Bob) onto one of the four maximally entangled Bell states,

\begin{eqnarray}
\ket{\psi^\pm_{AB}} =\frac{1}{\sqrt{2}}( \ket{e_A l_B} \pm  \ket{l_A e_B}), \nonumber \\
\ket{\phi^\pm_{AB}} =\frac{1}{\sqrt{2}}( \ket{e_A e_B} \pm  \ket{l_A l_B}).
\label{Bell}
\end{eqnarray}

Once a sufficiently large number of qubits has been transmitted to Charlie \cite{Curty2014}, he publicly announces which of his joint measurements resulted in one of these four states (photons often get lost during transmission, making a BSM impossible), and he also identifies the measurement result. This allows Alice and Bob to discard the records of qubits that did not generate a successful BSM. It is worth noting that to ensure security, Charlie only needs to be able to project onto one Bell state, but access to more Bell states will increase the key rate in MDI-QKD~\cite{Raju2014}. Next, based on the information from Charlie, Alice and Bob perform a basis reconciliation procedure known as key sifting. For every successful projection onto a Bell state at Charlie's, Alice and Bob use an authenticated public channel to reveal and compare the preparation bases for their respective qubits, X or Z, and keep only the record of events for which they have picked same basis. Depending on the result of the BSM and the users' preparation bases, Bob must post-process his bit values so that they become identical to Alice's. For instance, Bob performs a bit flip if the announced measurement resulted in $\ket{\psi^-_{AB}}$, which indicates anti-correlated bits in both the X- and Z-basis. 

Furthermore, due to imperfect qubit preparation, channel noise and noisy single-photon detectors, as well as possible eavesdropping, it is necessary to go through a key distillation process: (1) Alice and Bob publicly reveal a subset of their (non-discarded) bit values, and estimate the error rate for each basis independently; (2) they perform classical error correction on the bits resulting from Z-basis preparation; (3) the error rate for the X-basis is used to bound the information that an eavesdropper could have obtained during photon transmission and detection -- it is subsequently removed by means of privacy amplification. This results in a secure key, whose key rate is given by: 
\begin{equation}
S \geq [Q^Z[1-h_2(e^X)] - Q^Xf h_2(e^X)].
\label{keyrate_basic}
\end{equation}
Here, $Q^{Z(X)}$ refers to the gain, i.e. the probability of a projection onto a Bell state per emitted pair of qubits in either basis; $e^{Z(X)}$ denotes error rates, i.e. the ratio of erroneous to total projections onto a Bell state per emitted pair of qubits prepared either basis; $h_2$ is the binary Shannon entropy and $f\geq1$ characterizes the efficiency of error correction with respect to Shannon's noisy coding theorem.

The above-described protocol, which was originally proposed in 1996\cite{Biham96}, needs qubits encoded into genuine single photons. This is currently difficult to realize due to the lack of high-quality single photon sources. Fortunately, it is possible to overcome this problem by using phase-randomized attenuated laser pulses, which are easy to prepare using commercial technology, in conjunction with the so-called decoy state technique~\cite{Lo2012,Wang2013, Ma2012,Curty2014}. By randomly modulating the mean photon number of the laser pulses between several values known as `vacuum', `decoy' and  `signal'  \cite{Chan13,Xu2014}, one can assess lower and upper bounds on the gain and the error rate associated with Alice's and Bob's laser pulses both containing exactly one photon, respectively. In this case, the secret key rate is given by
\begin{equation}
S \geq [Q_{11}^Z[1-h_2(e^X_{11})] - Q^Z_{\mu\sigma}f h_2(e^Z_{\mu\sigma})].
\label{keyrate}
\end{equation}
The subscript `$11$' denotes values for gain and error rate stemming from Alice and Bob both emitting a single photon, and `$\mu\sigma$' denotes values associated with mean photon numbers per emitted pair of `signal' pulses of $\mu$ and $\sigma$, respectively. As in Eq. (\ref{keyrate_basic}), the basis is indicated by the superscript.


\section{Implementation of our MDI-QKD}
\begin{figure*}[htbp]
\begin{center}
{\includegraphics[width=\linewidth]{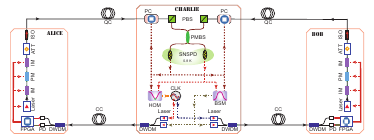}}
\caption{Experimental setup. PC: polarization controller, PBS: polarization beam splitter, PMBS: polarization maintaining beam splitter, SNSPD: superconducting nanowire single photon detector, HOM: Hong-Ou-Mandel measurement, CLK: clock, BSM: Bell state measurement, DWDM: dense wavelength division multiplexers, PD: photo-detector, FPGA: field-programmable gate array, IM: intensity modulator, PM: phase modulator, ATT: attenuator, ISO: isolator, QC: quantum channel, CC: classical channel. Note that the CLK and BSM signals are distributed to Alice and Bob electronically in the experiment.}
\label{set-up}
\end{center}
\end{figure*}

In this section, we describe the realization of our MDI-QKD system (a schematics is given in. Figure \ref{set-up}). We divide our system into four parts: qubit preparation modules, BSM module, control modules (ensuring indistinguishability of photons arriving at Charlie), and time-tagging module.  In the following subsections we describe the implementation of each module.

\subsection{Qubit preparation modules}
\label{sec:qubitprep}
As mentioned in section~\ref{sec:protocol}, Alice and Bob need to prepare qubits randomly in one of the four BB84 states. Furthermore, to use attenuated laser pulses as opposed to true single photons as carriers, they also have to vary the pulses' mean photon number and randomize the qubits' global phase. We employ, for both Alice and Bob, a distributed feedback (DFB) laser at 1548 nm wavelength combined with a home-built driver board. By operating the laser below and above the lasing threshold (i.e. operating it in gain-switching mode), we first generate phase-randomized laser pulses of 40 ns duration, which eliminates the possibility of an unambiguous-state-discrimination attack \cite{Tang2013,Dusek2000}. The pulses are then sent into an intensity modulator (IM) where, depending on the desired qubit state, early and/or late temporal modes of 200 ps duration and with 2.5 ns separation are carved out. The electrical pulses applied to the IM are created by an FPGA-based signal generator (SG). We feed the resulting optical signals into a phase modulator (PM) that can apply a phase shift of $\pi$ to the late mode, as determined by a random binary electronic signal created by the same SG. A second IM then allows to rapidly vary the overall intensity of these pulses, and a 99:1 beam-splitter combined with a photo-diode (not shown in Fig.~\ref{set-up}) is used in a feedback loop to maximize and maintain the extinction ratio of both IMs combined at about 60 dB. Finally, an optical attenuator allows reducing the intensity of all light pulses to the single photon level, and an optical isolator (ISO) with 50 dB isolation is used to shield Alice and Bob from Trojan horse attacks \cite{Gisin2006}. The current qubit generation rate is 20 MHz.  In our implementation, all electronic signals from the SG board are determined by binary random numbers that have been created off-line using a quantum random number generator (QRNG) \cite{Qiang2017}, and are stored in the FPGA. 

\subsection{Stabilization and feedback modules}
For Charlie to be able to perform a BSM successfully, the photons emitted by Alice and Bob need to be indistinguishable in all degrees of freedom, i.e spatial, spectral, polarization and temporal degrees. The spatial overlap is trivially guaranteed by the use of single mode fibers. The spectral overlap is ensured by carefully tuning and stabilizing the wavelengths of the DFB lasers used for creating qubits. On the other hand, since the photons generally travel long distances (tens of kilometres) through independent fibers, they are subjected to different time-varying environments. This results in fluctuating polarization states and arrival times at Charlie's. It is thus necessary to employ feedback mechanisms to actively compensate for these changes. For efficient key generation, it is desirable for the feedback systems not to interfere with the actual key distribution (ensuring maximum running time for key distribution), and for all the expensive components of the control module to be included into Charlie. This will allow several users in a future star-type network to share these resources, thereby adding to the cost-effectiveness of the network. As described in detail below, the feedback mechanisms used in our implementation satisfy these requirements.

\subsubsection{Spectral degree of freedom}
To ensure spectral overlap, we employ two continuous-wave DFB lasers with similar bandwidth. We tune their frequency with a resolution of 11.25 MHz by changing the temperatures of the laser diodes until the difference becomes small compared to the spectral width of the created (200 ps long) light pulses, which is 1.26 GHz assuming Gaussian shapes. Towards this end, we interfere unmodulated light emitted by the two lasers into extra fibers on a 50:50 beamsplitter and detect the beat signal (not shown in Fig.~\ref{set-up}). Using a proportional-integral-derivative (PID) feedback loop that acts on the laser temperature, the frequency difference remains below 20 MHz during a period of more than 40 hours. While it is possible to continuously monitor the frequency difference without affecting qubit generation, this was only necessary in the beginning of, but not during, a measurement. 
 
\subsubsection{Polarization degree of freedom}
To ensure that the photons coming from Alice and Bob have the same polarization at Charlie, we insert polarization beam splitters (PBSs) at the two inputs of the polarization maintaining beam splitter (PMBS) where the BSM takes place. Polarization fluctuations will thus be mapped onto fluctuations in the count rates of the two superconducting nanowire single photon detectors (SNSPDs) \cite{snspds} that are used to perform the BSM. These fluctuations control the settings of two polarization controllers (PC, General Photonics, Polastay-POS-002-E) that actively change polarization of input light until the single detector count rates are maximized and hence all polarization changes during photon transmission are compensated for. We emphasize that our method for polarization control does not require extra single photon detectors (SPDs) at Charlie, which differs from the method used in \cite{Tang200}. Moreover our approach results in continuously running polarization feedback, unlike other methods that necessitate interrupting the stream of quantum signals \cite{Wolfgang2013, Raju2015}.

\subsubsection{Timing}
To compensate for varying transmission times from Alice and Bob, respectively, to Charlie, we observe the degree of Hong-Ou-Mandel (HOM) quantum interference at Charlie~\cite{Raju2016}. To this end, the signals from his two SNSPDs are sent to a HOM unit (in addition to signalling projections onto Bell states, which is further described below), which monitors the rate of coincidence detections corresponding to either both photons arriving in mode $\ket{e}$, or both in mode $\ket{l}$. Thanks to the photon bunching, the coincidence count rate reaches a minimum when the photons from Alice and Bob arrive at the PMBS at exactly the same time, providing a precise feedback signal using which we keep arrival times locked. More precisely, we vary Alice's qubit generation time with a precision of 27.8 ps ($\sim$7.2 times less than the width of each temporal mode) to keep the coincidence count rate continuously at the minimum. Thus the arrival times are matched using a free-running feedback mechanism, which does not require the use of additional SNSPDs or high bandwidth PDs~\cite{Pan2016}.

\subsection{BSM module}
The BSM module for time-bin qubit-based MDI-QKD includes a PMBS followed by two SNSPDs, and a fast logical circuit triggered by the electrical signals from the two SNSPDs. For the two-photon projection measurement, the two indistinguishable photons -- one from Alice and one from Bob -- are overlapped at the PMBS to erase which-way information. A projection onto the $\ket{\psi^{-}_{AB}}$ Bell state is signalled by coincidence detection of two photons (one in each SNSPD) in orthogonal temporal modes (one in $\ket{e}$ and one in $\ket{l}$, while $\ket{\psi^{+}_{AB}}$ corresponds to coincidence detection of two photons in orthogonal temporal modes but in the same detector. It is worth noting that the BSM efficiency is limited to 50\% when only linear optics and no auxiliary photons are used \cite{Lütkenhaus1999}. In our implementation, we only select projections onto $\ket{\psi^{-}_{AB}}$ for secret key generation, which reduces the maximum efficiency (assuming lossless detection) to 25\%. The coincidence measurement is realized using a home-built broadband logical circuit with a coincidence window of $\sim$0.7 ns. 

\subsection{Time-tagging module}
For any MDI-QKD system, a time-tagging module is needed to record the information of Alice's and Bob's qubit preparations. In our case, this concerns the emission time with a precision of 50 ns plus four bits that specify the basis (X or Z), the bit value (0 or 1), and which out of three different mean photon numbers (vacuum, decoy, signal) have been chosen. Furthermore, the time of a successful BSM at Charlie plus the state projected onto (in our case only $\ket{\psi^-}$) must be registered. Knowing the exact travel times from Alice and Bob, respectively, to Charlie then allows back-tracking and establishing which two qubits have interacted at Charlie. Here we chose a simpler and less memory-intensive approach.   


In our MDI-QKD system, Charlie sends a common clock signal to synchronize the qubit preparation devices at Alice and Bob. During the time tagging process, Alice and Bob send the information of their prepared qubits (with the exception of time) into memory buffers, i.e. first-in-first-out (FIFO) buffers in their FPGAs, while the corresponding qubits are sent to Charlie. The memory buffer time is set to be equal to the time required by the qubits to reach Charlie plus the time required by the BSM signals to reach Alice (Bob) from Charlie. A simple logic operation then allows singling out only qubit generations that resulted in a successful BSM -- only those are further processed. 
   Note that our time-tagging module is integrated in the same FPGA that is also used as a SG (see section~\ref{sec:qubitprep}).

\section{Experimental results and discussion}

\begin{figure}[htbp]
\begin{center}
{\includegraphics[width=\linewidth]{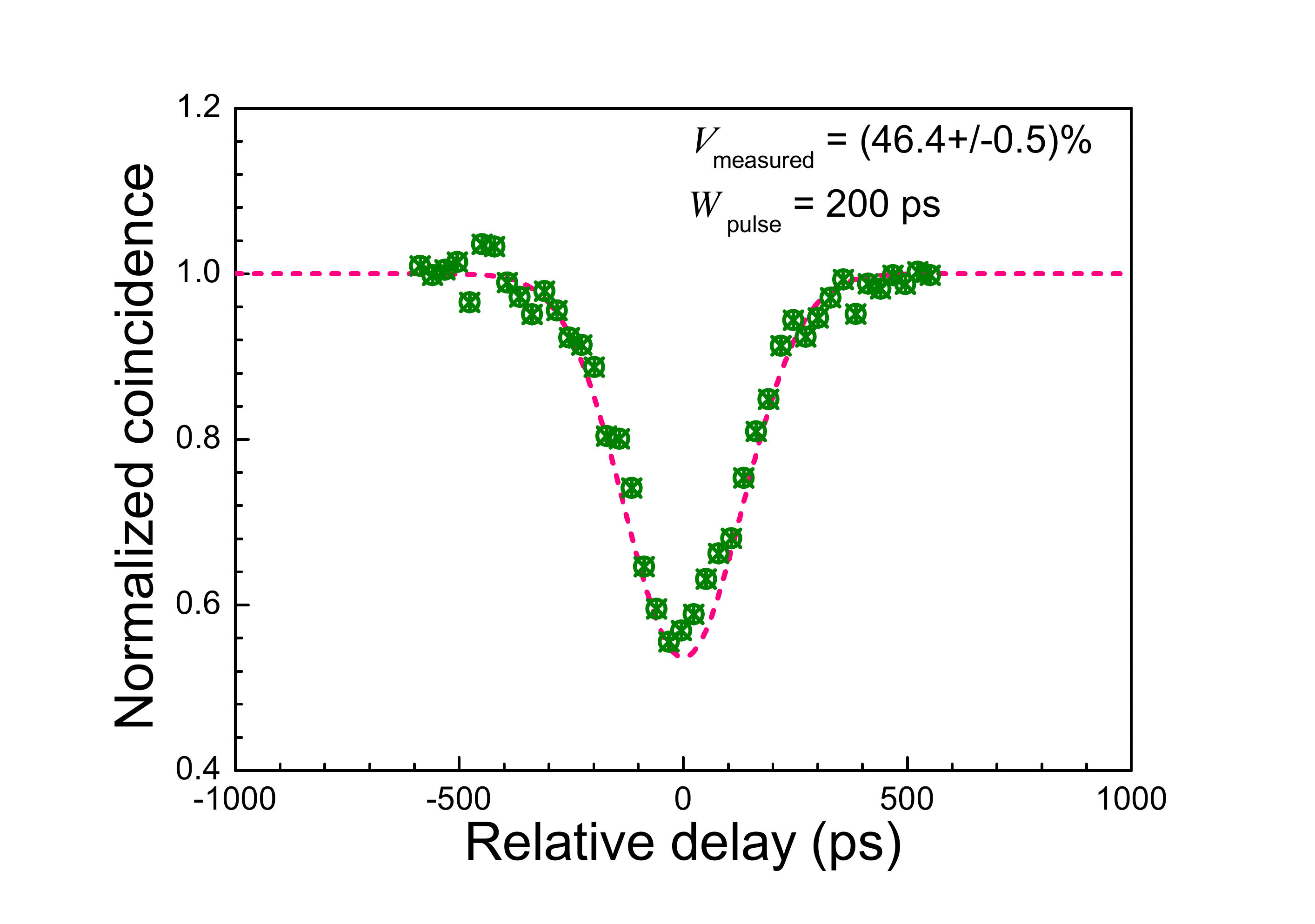}}
\caption{Result of HOM interference between two independent qubits, encoded into phase-randomized attenuated laser pulses, after each has traveled over 40 km of spooled fiber. Dots show the measured results; the dashed line is the simulation with 200 ps long Gaussian pulses. The maximally achievable visibillity is 50\%.}
\label{hom}
\end{center}
\end{figure}

We first test the indistinguishability between the photons from Alice and Bob by measuring the visibility of HOM interference using two spooled fibres of 40 km length. Figure~\ref{hom} shows the result; the mean photon number is 0.03 per qubit. The circles are the experimentally measured values, while the dashed line is a fit assuming 200~ps wide Gaussian pulses. A visibility of 46.4$\pm$0.5\% is obtained in our measurement, which is slightly smaller than the maximally possible value of 50\% for pulses with Poissonian photon-number distribution. The difference is due to residual distinguishability of the two photons, such as residual difference of spectral and temporal modes.

\begin{figure}[htbp]
\begin{center}
{\includegraphics[width=\linewidth]{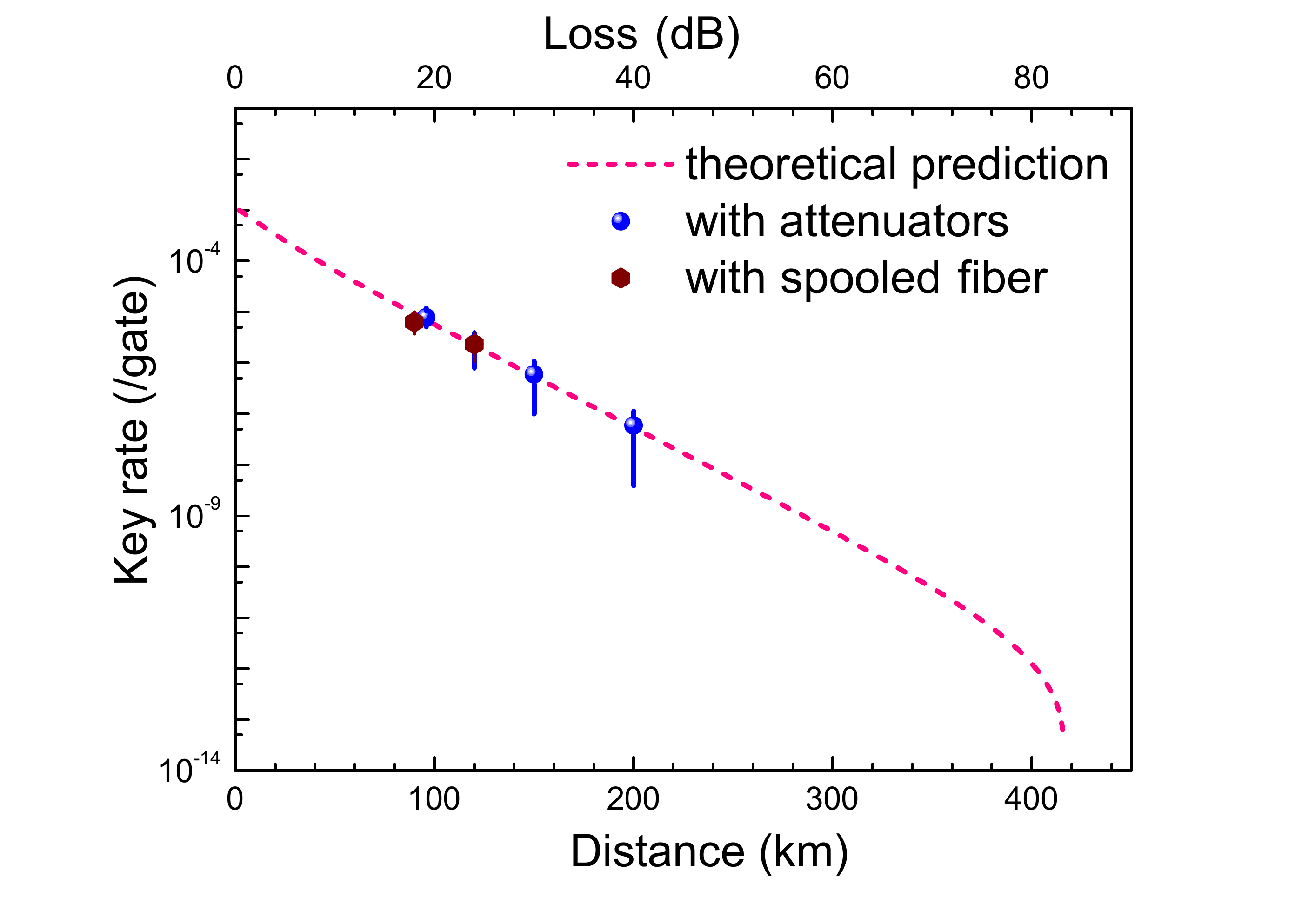}}
\caption{Secure key rates. Circles denote measured results with attenuators simulating loss in the quantum channels, and diamonds represent measurements with spooled fiber. }
\label{key}
\end{center}
\end{figure}

Next we run the full QKD system and assess its performance. In the experiment, Alice and Bob prepare time-bin qubits based on pre-stored random numbers and send them to Charlie. Once Charlie successfully projects onto the $\ket{\psi^-}$ Bell state, he sends a signal to Alice and Bob, who tag the corresponding random numbers (i.e. qubit states) on their corresponding time-tagging modules. After accumulating about 30 million bytes of tagged data, Alice and Bob compare their files. Based on the gains and quantum bit error rates (QBERs) obtained from this comparison, we calculate the achievable secure key rate \cite{Chan13}. Figure~\ref{key} shows the experimental results and the theoretical prediction for the different setups used in our demonstrations. The performance is tested with spooled fibre of 2$\times$40 km and 2$\times$60 km (diamonds), and with attenuators simulating lossy quantum channels (circles). We emphasize that all feedback mechanisms are running continuously during the measurements. We find that all measurement results agree well with the prediction (the dashed curve). In particular, the secret key rate over 80 km of spooled fiber is of about 0.1 kbps, which can be improved to around 10 kbps by increasing the clock rate from the current 20 MHz to 2 GHz -- the maximum allowed by the 200 ps time-jitter of the SNSPDs \cite{Raju2014, snspds}. Finally, we note that the performance of our implementation predicts positive secure key rates over up to 400 km assuming (standard) fiber with 0.2 dB loss/km. This distance can be increase to 500 km when using ultra low-loss fibre (0.16 dB loss/km) as in \cite{Yin404, Hugo2015}.

\section{Conclusion and outlook}
We have demonstrated a MDI-QKD system based on cost-effective hardware such as  commercial DFB lasers;  FPGA-based  preparation, time-tagging and timing control;  and active feedback control for frequency, polarization and arrival-time of  photons. Our demonstration paves the way for MDI-QKD-based star-type quantum networks with kbps secret key rates spanning geographical distances in excess of 100 km. 

\section*{Acknowledgments}

We acknowledge technical support by Vladimir Kiselyov. This work was supported through Alberta Innovates Technology Futures, the National Science and Engineering Research Council of Canada (through their Discovery Grant and the CryptoWorks 21 CREATE programs), the Calgary Urban Alliance and the US Defense Advanced Research Projects Agency InPho Program. Part of the research was carried out at the Jet Propulsion Laboratory, California Institute of Technology, under a contract with the National Aeronautics and Space Administration. W.T. is a Senior Fellow of the Canadian Institute for Advanced Research.

\ifCLASSOPTIONcaptionsoff
  \newpage
\fi

\end{document}